\documentclass[]{iopart}
\usepackage{graphicx,epsf, epsfig, amssymb}
\usepackage[usenames]{color}

\newcommand{\be}{\begin{equation}}
\newcommand{\ee}{\end{equation}}
\newcommand{\bel}[1]{\begin{equation}\label{#1}}
\newcommand{\ba}{\begin{eqnarray}}
\newcommand{\ea}{\end{eqnarray}}
\newcommand{\bal}[1]{\begin{eqnarray}\label{#1}}

\newcommand{\url}[1]{{\it #1}}

\begin{document}

\title{Can we Detect Intermediate Mass Ratio Inspirals?}
\author{Ilya Mandel$\footnote[1]{email: ilyamandel@chgk.info}$}
\address{Department of Physics and Astronomy, Northwestern University, Evanston, IL, 60208 USA}
\author{Jonathan R.~Gair$\footnote[2]{email: jgair@ast.cam.ac.uk }$}
\address{Institute of Astronomy, University of Cambridge, Cambridge, CB30HA, UK}

\begin{abstract}
Gravitational waves emitted during intermediate-mass-ratio inspirals
(IMRIs) of intermediate-mass black holes (IMBHs) into supermassive
black holes could represent a very interesting source for LISA.
Similarly, IMRIs of stellar-mass compact objects into IMBHs could be
detectable by Advanced LIGO.  At present, however, it is not clear
what waveforms could be used for IMRI detection, since the
post-Newtonian approximation breaks down as an IMRI approaches the
innermost stable circular orbit, and perturbative solutions are
only known to the lowest order in the mass ratio.  We discuss the
expected mismatches between approximate and true waveforms, and the
choice of the best available waveform as a function of the mass ratio
and the total mass of the system.  We also comment on the significance
of the spin of the smaller body and the need for its inclusion in the
waveforms.

      %We also comment on the systematic errors in parameter estimation that will
      %arise due  to the use of approximate IMRI waveforms.

\end{abstract}

\maketitle

\section{Introduction}

Observational evidence from cluster dynamics and from ultra-luminous X-ray sources 
suggests that there may exist a population of intermediate-mass black holes (IMBHs) with 
masses in the M $\sim 10^2$ -- $10^4 M_\odot$  range \cite{MillerColbert:2004, Trenti:2006, Noyola:2008}. Numerical simulations of globular clusters suggest that IMBHs could merge with numerous lower-mass compact objects (COs) during the lifetime of the cluster \cite{Taniguchi:2000, MillerHamilton:2002a, MillerHamilton:2002b, MouriTaniguchi:2002a, MouriTaniguchi:2002b, Gultekin:2004, Gultekin:2006, OLeary:2006},  through a combination of emission of gravitational radiation, binary exchange processes, and secular evolution of hierarchical triple systems.   The evidence for the existence of IMBHs is still inconclusive, however.  There is much debate over the formation of IMBHs, and some of the evidence cited in favor of IMBHs could have alternative explanations \cite{Hurley:2007, Colbert:2008}.   We refer the reader to \cite{MillerColbert:2004, Miller:2009} for thorough reviews.

If IMBHs exist, gravitational waves (GWs) generated during mergers of IMBHs with other compact objects are potentially detectable by existing or proposed GW detectors.  In this paper, we focus on GWs generated during intermediate-mass-ratio inspirals (IMRIs).  Rate predictions for such events are extremely uncertain at present, but several mechanisms have been proposed that {\it may} lead to detectable IMRIs.  The next generation of the ground-based gravitational-wave detector, the Laser Interferometer Gravitational-Wave Observatory (Advanced LIGO \cite{Smith:2009}), could detect the GWs generated during an IMRI of a stellar-mass object (black hole or neutron star, since a white dwarf or a main sequence star would be tidally disrupted) into an IMBH of mass $\lesssim 350 M_\odot$ at a rate of a few per year \cite{Mandel:2007rates}.   Meanwhile, if a cluster containing an IMBH starts out sufficiently close to a supermassive black hole (SMBH) residing in a galactic center, it will sink to the center and, after releasing the IMBH due to tidal stripping of the cluster \cite{Ebisuzaki:2001}, will form an IMBH-SMBH binary \cite{Miller:2005}.  This binary will eventually inspiral under the influence of radiation reaction, creating an IMRI whose frequencies will make it detectable by a proposed  space-based detector, the Laser Interferometer Space Antenna (LISA \cite{Bender:1998}) \cite{Amaro:2007}; LISA could detect a few such events per year \cite{Miller:2005}.  LISA could also detect inspirals of white dwarfs into black holes in the $10^4$ -- $10^5\ M_\odot$ range with possible electromagnetic counterparts due to white-dwarf tidal disruption \cite{Sesana:2008}.

We emphasize that predictions of possible rates of detections of GWs from IMRIs are highly uncertain, and depend on a number of debated assumptions about IMBH formation mechanisms, dynamics of globular clusters or galactic nuclei, etc.  Nonetheless, the scientific potential of IMRIs involving IMBHs is sufficient to make them interesting candidates for LIGO and LISA searches.  A detection of an IMRI could provide the first confirmation of IMBH existence in the absence of definitive electromagnetic observations.  Further detections of LIGO IMRIs would make it possible to explore the dynamics of globular clusters, while LISA IMRIs will additionally elucidate the dynamical processes in galactic nuclei.    However, detection and especially parameter estimation of IMRIs will require the construction of accurate IMRI waveforms.  

In this paper we analyze the suitability of currently available waveforms for LISA IMRI detection. Currently, two different types of approximate inspiral waveforms are available.  On the one hand, there are post-Newtonian (pN) waveforms, which are expansions in the velocity $v/c$ (\cite{Blanchet:2006} and references therein).  On the other hand, there are perturbative waveforms for extreme-mass-ratio inspirals (EMRIs) \cite{BarackCutler:2004, Sago:2008} which are expansions in the dimensionless mass ratio $\eta\equiv M_1 M_2/(M_1+M_2)^2$.  Post-Newtonian waveform are known to 3.5pN order ($(v/c)^7$), but the convergence of this series deteriorates at the high velocities reached near the innermost stable circular orbit where an IMRI spends a significant number of cycles (this number of cycles scales as one over the mass ratio).  Meanwhile, EMRI waveforms are only known to the lowest order in the mass ratio \cite{Poisson:2004,Gralla:2008}, and their accuracy deteriorates at intermediate mass ratios. The first question one might ask is: at what value of $\eta$ does the EMRI waveform become more faithful than the pN waveform? An order-of-magnitude approach to an answer might proceed as follows.  For a moderate value of the spin of the more massive body, say, $\chi \equiv S_1/M_1^2 \equiv a_1/M_1 \sim 0.5$, the comparable-mass case is characterized by the orbital angular momentum of the binary at the innermost stable circular orbit (ISCO) dominating over the spin angular momentum, while in the extreme-mass-ratio inspiral the spin angular momentum of the massive body dominates over the orbital angular momentum of the binary.  For $\chi=0.5$, this transition (as computed for a Keplerian orbit at the ISCO) occurs at a mass ratio of approximately $5:1$. But does this necessarily mean that intermediate mass ratios of $10:1$ or $100:1$ or even $1000:1$ are faithfully represented by EMRI orbits? In this paper we demonstrate that there is a significant range of intermediate mass ratios over which neither the pN nor the EMRI waveforms are likely to be faithful, and advocate the urgent need to develop more suitable IMRI waveforms.

The paper is organized as follows.  In Section \ref{sec:pNvsEMRI}, we estimate and compare the errors inherent in the post-Newtonian and extreme-mass-ratio approximations.  In Section \ref{sec:spin}, we analyze the impact of the spin of the smaller object on the waveform and the need to include spin-spin coupling for the purposes of signal detection.  Finally, in Section \ref{sec:future}, we discuss the future prospects for developing an accurate family of IMRI waveforms.

\section{Waveform comparison: pN vs.~EMRI}\label{sec:pNvsEMRI}

\subsection{Waveforms}

To determine the faithfulness of the two approximate waveform families, we should compare each with the true theoretical waveform, ``nature's waveform''.  Unfortunately, we do not have such a waveform at our disposal.  Despite the remarkable recent advances of numerical relativity (\cite{Pretorius:2005, Scheel:2008} and many others), it remains unlikely that long numerical simulations of inspirals with mass ratios of $100:1$ or $1000:1$ will be available in the foreseeable future.  In the absence of a true waveform to compare against, we can estimate the range of validity of the two approximate waveform families as follows.  

To test the range of validity of the post-Newtonian waveform, we compare the 3.5 pN waveform with the 3 pN waveform, following the example of \cite{CutlerVallisneri:2007}.  In the range of validity, the difference between these two waveforms, i.e., the 3.5 pN term in the expansion, should serve as a reasonable estimate for the difference between the full waveform and the approximate post-Newtonian expansion.  The point at which the overlap between these waveforms drops significantly from 1 signifies the end of the range of validity of the post-Newtonian approximation.   

We can not use a similar trick for EMRI waveforms, since they are not yet available at the second order, despite significant recent progress~\cite{Poisson:2004,Sago:2008,Gralla:2008}.  Instead, we estimate the range of validity of the perturbative approximation that ignores higher-order terms in the mass ratio by comparing the full 3.5 pN waveform with a 3.5 pN waveform that includes only the lowest-order terms in the mass ratio.  As the mass ratio is increased, the overlap between the full and EMRI-fied 3.5 pN waveforms may decrease significantly from 1; this is the mass ratio at which the EMRI approximation ceases to be reliable.

The time-domain post-Newtonian waveform is given by
\be
h(t)=A(t) e^{i \Phi};
\ee
the time-domain phase $\Phi$ can be expressed in terms of the GW frequency $f\equiv\ 1/(2\pi)\ 
d\Phi/dt$ (multiplying Eq.~(235) of \cite{Blanchet:2006} by $2$ to get the GW phase from the orbital phase):
\ba
\Phi &=& -\frac{x^{-5/2}}{16 \eta} \Biggl\{1+\left(\frac{3715}{1008}+\alpha\frac{55}{12} \eta\right)x-10\pi x^{3/2}\\
	\nonumber
	&+&
    \left(\frac{15293365}{1016064}+\alpha\frac{27145}{1008} \eta+ \alpha\frac{3085}{144} \eta^2\right) x^2\\
    \nonumber
    &+&
    \left(\frac{38645}{1344}-\alpha\frac{65}{14}\eta\right)\pi x^{5/2}\log \frac{x}{x_0}\\
    \nonumber
    	&+&
   \biggl[\frac{12348611926451}{18776862720}-\frac{160}{3}\pi^2-\frac{1712}{21} C-\frac{856}{21} \log(16 x)\\ 
    \nonumber
    & &+ \alpha\left(-\frac{15737765635}{12192768}+\frac{2255}{48}\pi^2\right)\eta + \alpha\frac{76055}{6912}\eta^2-\alpha\frac{127825}{5184}\eta^3\biggr] x^3\\
    \nonumber
    &+&
    \beta \left(\frac{77096675}{2032128}+\alpha\frac{378515}{12096}\eta-\alpha\frac{74045}{6048}\eta^2\right)\pi x^{7/2} \Biggr\},
\ea
where $M=M_1+M_2$ is the total mass, $C\approx 0.577$ is Euler's constant, $x\equiv(\pi M f G c^{-3})^{2/3}$, and $x_0=x(f=0.1{\ \rm{mHz}})$.  We introduced the parameters $\alpha$ and $\beta$ to distinguish the waveform families of interest.  The {\it ``base''} waveform uses the full 3.5 pN phase: 
\be
\Phi_{\rm base}=\Phi(\alpha=1,\ \beta=1).
\ee
The approximate  {\it post-Newtonian} waveform has the 3 pN phase: 
\be \Phi_{\rm pN}=\Phi(\alpha=1,\ \beta=0).
\ee
Finally, the approximate {\it EMRI} waveform has the 3.5 pN phase expanded to the lowest order in $\eta$ only: 
\be
\Phi_{\rm pN}=\Phi(\alpha=0,\ \beta=1).
\ee
 
To compute overlaps of waveforms weighted by the noise power spectral density of the LISA detector, as described below, it is more convenient to use the frequency-domain representation of the waveform
\be
\tilde{h}(f)=A(f) e^{i \psi}.
\ee
For simplicity, the restricted post-Newtonian waveform is computed via the stationary-phase approximation (e.g., \cite{PoissonWill:1995}), and the lowest-order Newtonian amplitude is used:
\be
A(f)=\frac{2}{D} \left(\frac{5 \mu}{96}\right)^{1/2} \left(\pi^2 M\right)^{1/3} f^{-7/6},
\ee
where $D$ is the distance to the source and $\mu=M_1 M_2 / M$.   The expression for the phase $\psi(f)$ is (e.g., Eq.~(3.4) of \cite{Arun:2005}):
\ba
\psi &=& 2\pi f t_c - \phi_c - \frac{\pi}{4}+ \frac{3 x^{-5/2}}{128 \eta} \Biggl\{1+\frac{20}{9}\left(\frac{743}{336}+\alpha\frac{11}{4} \eta\right)x\\
	\nonumber
	&-&16\pi x^{3/2}
    +10\left(\frac{3058673}{1016064}+\alpha\frac{5429}{1008} \eta+ \alpha\frac{617}{144} \eta^2\right) x^2\\
    \nonumber
    &+&
    \pi \left(\frac{38645}{252}-\alpha\frac{65}{3}\eta\right)\pi x^{5/2}\log \frac{x}{x_0}\\
    \nonumber
    	&+&
   \biggl[\frac{11583231236531}{4694215680}-\frac{640}{3}\pi^2-\frac{6848}{21} C-\frac{6848}{21} \log(4 x)\\ 
    \nonumber
    & &+ \alpha\left(-\frac{15335597827}{3048192}+\frac{2255}{12}\pi^2+\frac{1760}{3}\frac{11831}{9240} -\frac{12320}{9}\frac{1987}{3080}\right)\eta\\
    \nonumber
    & &  + \alpha\frac{76055}{1728}\eta^2-\alpha\frac{127825}{1296}\eta^3\biggr] x^3\\
    \nonumber
    &+&
    \beta\left(\frac{77096675}{254016}+\alpha\frac{378515}{1512}\eta-\alpha\frac{74045}{756}\eta^2\right)\pi x^{7/2} \Biggr\},
\ea
where $t_c$ is the time of coalescence (the time where the frequency would formally go to infinity), and $\phi_c$ is the phase at coalescence.  Again, setting $\alpha=\beta=1$ corresponds to $\psi_{\rm base}$, setting $\alpha$ to 0 yields $\psi_{\rm EMRI}$, and $\psi_{\rm pN} = \psi(\alpha=1,\ \beta=0)$

We use the lowest-order (0 PN) relationship between time and frequency to determine the starting frequency for a year-long inspiral signal.  We define the dimensionless time variable $\Theta$ as
\be
\Theta \equiv \frac{\eta c^3}{5 G M} (t_c - t),
\ee
where $t_c$ is the time of coalescence.   Then at 0 PN
\be
\Theta \approx \frac{1}{256} x^{-4} = \left(\frac{8 \pi G M f}{c^3}\right)^{-8/3}.
\ee

\subsection{Comparison}

\begin{figure}[htb]
\begin{center}
\includegraphics[keepaspectratio=true, width=\textwidth]{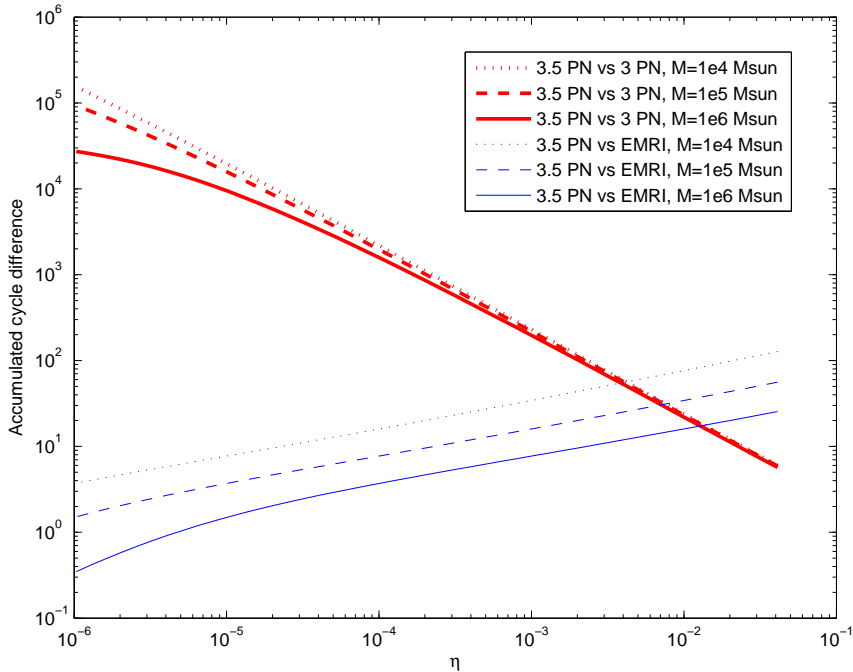}
\end{center}
\caption{The difference in cycles between approximate and base waveforms over the last year of inspiral before ISCO (or during the time it takes the GW frequency to increase from $0.01$ mHz to the ISCO, if less), as a function of the symmetric mass ratio $\eta$. Thick (red) curves show  $(\Phi_{\rm base}-\Phi_{\rm pN})/(2\pi)$; thin (blue) curves show $(\Phi_{\rm base}-\Phi_{\rm EMRI})/(2\pi)$.  Solid, dashed, and dotted curves refer to inspirals into Schwarzschild black holes of $10^6$, $10^5$, and $10^4$ solar masses, respectively.  \label{fig:LISAcycles}}
\end{figure}

We begin the comparison of the approximate waveforms with the ``base'' waveform by measuring the number of cycles of difference accumulated over the last year of inspiral observed by LISA before the inspiraling object reaches the ISCO (or during the time it takes for the inspiral to proceed from a gravitational-wave frequency of $0.01$ mHz to ISCO, if that is less than a year).  In Fig.~\ref{fig:LISAcycles}, we plot the accumulated cycles of difference for the pN waveform, $(\Phi_{\rm base}-\Phi_{\rm pN})/(2\pi)$, and the EMRI waveform, $(\Phi_{\rm base}-\Phi_{\rm EMRI})/(2\pi)$, as a function of $\eta$ for several choices of the mass of the central Schwarzschild black hole.  

We expect post-Newtonian excess cycles to accumulate near the end of the inspiral, where $v/c$ becomes significant ($v/c \sim 0.41$ near the Schwarzschild ISCO), so that the high-order pN terms have a sizable contribution.  

On the other hand, EMRI excess cycles should be more evenly spread throughout the inspiral.  The error in the lowest-order EMRI approximation results in an error in the frequency evolution of order 
\be
\delta \dot{f} = O(\eta^2),
\ee
while the  frequency evolution itself scales as
\be
\dot{f} \equiv \frac{df}{dt} = O(\eta).
\ee
The accumulated excess cycles thus scale as
\be
\delta \Phi_{\rm EMRI} = O((\delta \dot{f}) T^2) = O(\eta^2) T^2,
\ee
where $T$ is the time observation.  When the time of observation is limited by the duration of the LISA observation window, e.g., $T=1$ year, the accumulated excess cycles in the EMRI waveform scale as $O(\eta^2)$.  This is indeed the case for extreme mass ratios $\eta \ll 1$, which explains why even lowest-order EMRI waveforms may be sufficient for LISA EMRI detections.  However, for more rapidly evolving IMRIs, the observation time is limited by the fixed LISA bandwidth $\Delta f$:
\be
T \sim \frac{\Delta f}{\dot{f}} =(\Delta f) O(\eta^{-1}).
\ee
Then the accumulated excess cycles actually scale as $O(1)$, so we do not expect EMRI waveforms to produce a good match in this regime.

Indeed, these are precisely the trends that we observe in Fig.~\ref{fig:LISAcycles}.  Post-Newtonian waveforms accumulate few cycles of difference for large $\eta$, where the inspiral proceeds rapidly through the domain of high $v/c$, but they accumulate huge excess cycles at low $\eta$, when the inspiraling object spends $O(1/\eta)$ cycles near ISCO.  EMRI waveforms, meanwhile, accumulate $O(1)$ excess cycles when $\eta$ is large enough for the duration of observation to be limited by the LISA bandwidth rather than by the $1$-year observation window.

Comparing 3.5 PN to 3 PN, we find that there are more cycles of difference at higher masses for a given $\eta$. This is because higher mass systems spend proportionally longer at higher speeds $v/c$. Mathematically, this occurs because  the 3.5 PN term goes as $\theta^{-1/4}$, and $\theta \propto 1/M$, so the difference scales positively with mass.  When comparing EMRI-fied 3.5 PN and 3.5PN, we recall that the cycle differences are accumulated fairly uniformly over the course of the inspiral; since lower-mass systems undergo more cycles of oscillation in total over one year, such systems show the greater phase difference for a given value of $\eta$.

We note that we have included points in Fig.~\ref{fig:LISAcycles} where the phase difference is considerably larger than one cycle. In this regime, the approximation has broken down and so we clearly can not regard $\Phi_{base}$ to be the ``true'' waveform. However, we include the full range of mass ratios in the figure to more clearly illustrate the wide range of mass ratios where neither of our template families can be applied for the full duration of the inspiral.

%In the comparison of EMRI-fied 3.5 PN and 3.5PN there is a competition between two effects --- the fact that higher-mass systems spend longer at larger values of $v/c$, and the fact that lower-mass systems undergo more cycles of oscillation in total over one year. The higher mass ratio corrections enter at all PN orders, but become more important at higher $v/c$. At low values of $\eta$, the first factor wins and higher-mass systems show more cycles of phase difference, while when $\eta \to 0.25$ the latter factor wins and lower-mass systems show the greater phase difference.

Although comparing the number of excess cycles provides a good indication of the regime of validity of the two approximations, a more precise comparison should take the frequency-dependent instrumental noise into account: after all, if the excess cycles are accumulated at frequencies at which LISA is insensitive, they may not cause a significant issue for IMRI detections.  We define the overlap of two waveforms $\tilde{h}(f)$ and $\tilde{g}(f)$ as
\bel{eq:overlap}
\langle h|g \rangle = 4 \Re \int_{\rm flow}^{\rm fISCO} \frac{\tilde{h}(f) \tilde{g}^*(f)}{S_n(f)} df,
\ee
where $\tilde{g}^*$ denotes the complex conjugate of $\tilde{g}$, and $S_n(f)$ is the noise power spectral density of LISA, which includes both the instrumental noise and the unresolvable foreground of Galactic white dwarf binaries. We took our prescription for the LISA sensitivity curve and the white dwarf confusion foreground from~\cite{BarackCutler:2004}. We then define the normalized match of these two waveforms, ${\cal M}(h,g)$, as
\be
{\cal M}(h,g) = \frac{\langle h | g \rangle}{\sqrt{ \langle h | h \rangle  \langle g | g \rangle  }}.
\ee

\begin{figure}[htb]
\begin{center}
\includegraphics[keepaspectratio=true, width=\textwidth]{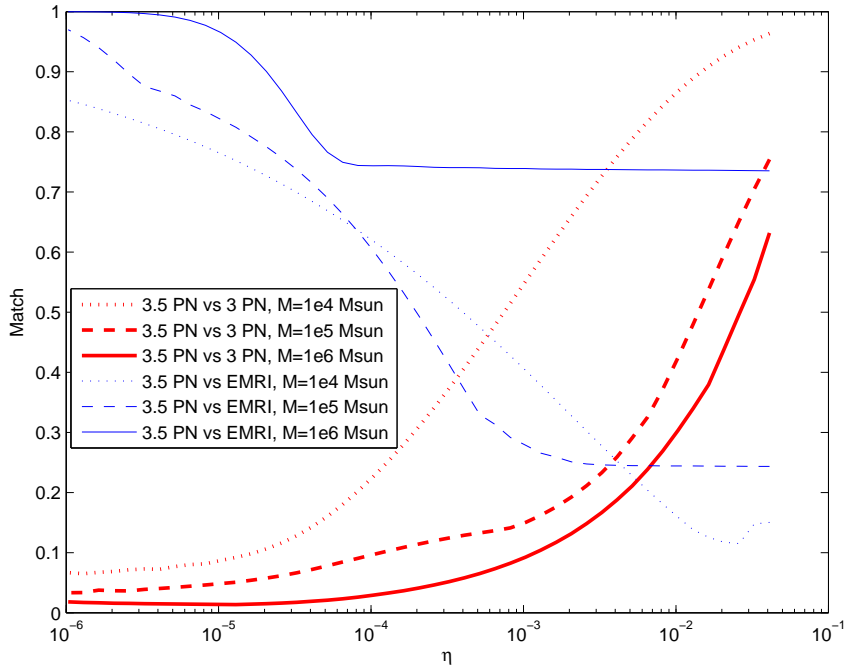}
\caption{The match factors between approximate and base waveforms over the last year of inspiral before ISCO (or during the time it takes the GW frequency to increase from $0.01$ mHz to the ISCO, if less), as a function of $\eta$. Thick (red) curves show  ${\cal M}(h_{\rm base}, h_{\rm pN})$;  thin (blue) curves show  ${\cal M}(h_{\rm base}, h_{\rm EMRI})$.  Solid, dashed, and dotted curves refer to inspirals into Schwarzschild black holes of $10^6$, $10^5$, and $10^4$ solar masses, respectively.  \label{fig:LISAoverlap}}
\end{center}
\end{figure}

In Fig.~\ref{fig:LISAoverlap}, we plot the matches between pN and EMRI waveforms on the one hand and the ``base'' waveform on the other.  We automatically maximize over possible constant phase shifts (different values of $\phi_c$) between two waveforms by taking the absolute value of the overlap integral instead of the real part in Eq.~(\ref{eq:overlap}).  We also maximize over different values of the time of coalescence, $t_c$.  Because the way in which the parameter $t_c$ enters Eq.~(\ref{eq:overlap}) essentially corresponds to a Fourier transform, it is relatively computationally inexpensive to do so.   We do not, however, maximize over different choices of intrinsic parameters, such as the masses of the two objects.  

We find, again, that EMRI waveforms are more faithful at low $\eta$ while pN waveforms are more faithful at high $\eta$.  For a given value of $\eta$, pN waveforms are more generally faithful for a lower central black-hole mass.  This is because the frequency at ISCO scales as $O(1/M)$, so for lower masses, the ISCO frequency is too high for LISA to have significant sensitivity there; instead, most of the signal-to-noise ratio is contributed by earlier parts of the inspiral, where $v/c$ is lower and the post-Newtonian approximation is still valid.

The most striking feature of Fig.~\ref{fig:LISAoverlap}, however, is that for a wide range of intermediate mass ratios, $10^{-5} \lesssim \eta \lesssim 10^{-1}$ (depending on total mass), neither the post-Newtonian nor the extreme-mass-ratio waveforms appear to be valid.  It is worth pointing out, however, that we have not maximized the match factors over the masses, only over the extrinsic parameters $t_c$ and $\phi_c$.  Therefore, either EMRI or pN waveforms could still be an {\it effective} waveform family in the sense that they could densely cover the parameter space of true waveforms without necessarily matching the true waveforms well for identical parameter values.  

\section{Small-body spin}\label{sec:spin}

An interesting additional question concerns the significance of the spin of the small body.   Earlier treatments of EMRI waveforms have generally ignored the spin of the smaller body (e.g., \cite{BarackCutler:2004, MLDC2}).  Indeed, for EMRIs, the dominant effect of the spin of the smaller body comes through the spin-spin coupling term, which enters the post-Newtonian expansion at 2 pN order \cite{Kidder:1993}.  This term is considerably smaller, especially for low $\eta$, than the coupling term between the spin of the large body and the orbital angular momentum, which enters at the 1.5 pN order.  Here, we use the same formalism as above to check whether we are indeed justified in neglecting the spin of the small body for EMRIs and IMRIs.

\begin{figure}[htb]
\begin{center}
\includegraphics[keepaspectratio=true, width=\textwidth]{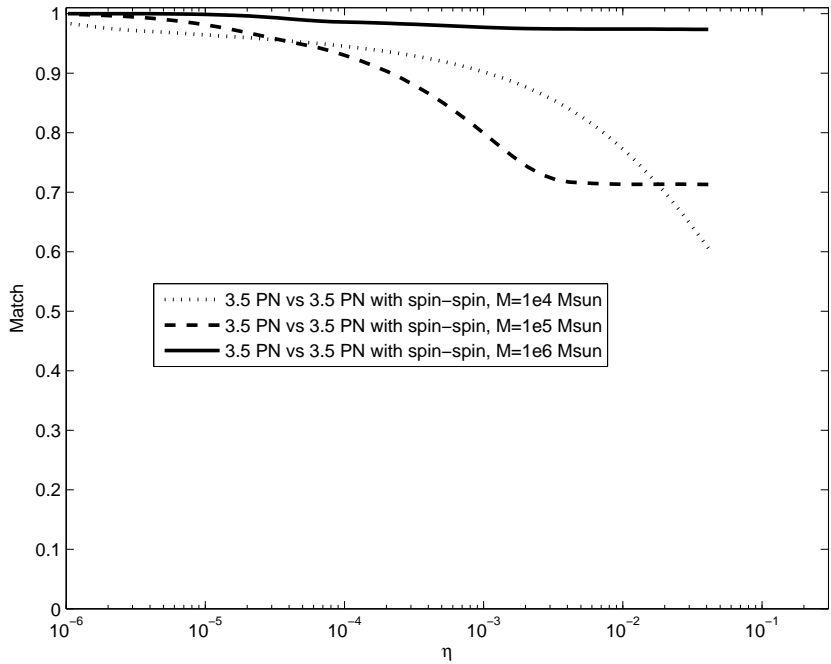}
\caption{Match factors ${\cal M}(h_{\rm base}, h_{\rm spin})$ between waveforms that include and omit spin-spin coupling, computed over the last year of inspiral before ISCO (or during the time it takes the GW frequency to increase from $0.01$ mHz to the ISCO, if less), as a function of  $\eta$.  Solid, dashed, and dotted curves refer to inspirals into Schwarzschild black holes of $10^6$, $10^5$, and $10^4$ solar masses, respectively. \label{fig:LISAspin}}
\end{center}
\end{figure}

We model the spin-spin coupling by adding a term corresponding to the maximal possible value of this coupling to the standard 3.5 pN phase (see Eqs.~(1.5), (3.2), and (3.6) of \cite{PoissonWill:1995}):
\be
\Phi_{\rm spin} = \Phi_{\rm base} - 5 \sigma x^2; \qquad \psi_{\rm spin} = \psi_{\rm base} - 10 \sigma x^2,
\ee
where 
\be
\sigma = (721-247)/48 \eta.
\ee  
We are not being consistent here, since the original waveform $\Phi_{\rm base}$ corresponds to two non-spinning objects, whereas the spin-spin term we have added corresponds to two objects that are maximally spinning and optimally aligned to produce the largest possible value of spin-spin coupling.  However, our purpose here is not to create an accurate model of a spinning binary, but rather to estimate the effect of ignoring the spin of the smaller body when the spin-spin coupling is present.  We do that, as in the previous section, by computing the match factor between $h_{\rm spin}$ and $h_{\rm base}$, which we plot in Fig.~\ref{fig:LISAspin} for various values of the central body's mass.

As expected, the spin-spin coupling is weak at extreme mass ratios.  We note that spin-spin coupling becomes important at masses ratios as low as $\sim 10^{-4}$ -- $10^{-3}$ for low-mass central black holes. This is a much more extreme mass ratio than usually supposed, so ignoring the spin of the smaller body would risk significantly reducing the match factor, and thus the detection efficiency.  The effect of the spin-spin term is suppressed for high total mass because the inspiral occurs at a lower frequency, so the amount of dephasing due to spin is limited by the small number of inspiral cycles in band; when $M=10^6\ M_\odot$, the match does not drop below $0.97$.

%Generally, the drop from near-unity match occurs at higher $\eta$ for lower masses; this is because for low total mass most of the LISA signal-to-noise ratio is contributed by the fraction of the inspiral long before ISCO, where even the 2 pN spin-spin coupling term is not very significant.  However, for the higher mass $M=10^6 M_\odot$, the overlap again increases as $\eta \to 0.25$, because the inspiral occurs at lower frequency, so the amount of dephasing due to spin is limited by the small number of inspiral cycles in band.

\section{Conclusions and future directions}\label{sec:future}

We have shown in this paper that there exists a significant problem with IMRI waveforms: for a wide range of intermediate mass ratios, both of the presently available approximate waveform families (post-Newtonian or EMRI) appear to be invalid, indicating the need for the development of a new waveform family for intermediate-mass-ratio inspirals.  Specifically, if we place a limit at a match factor of $0.9$, which would correspond roughly to a 30\% loss in detection efficiency, we find that both the post-Newtonian and perturbative approximations are invalid between mass ratios of $\sim 2\times 10^{-5}$ and $10^{-2}$ for all SMBH masses that we considered.   We point out, however, that we have maximized the overlap over the phase and time of coalescence only, so our results may underestimate the extent to which these approximate waveforms cover the space of ``true'' waveforms once variation in the intrinsic parameters (masses) is allowed.

We have also found that spin-spin coupling can not be ignored for mass ratios $\eta\gtrsim5\times10^{-3}$ for total mass $M=10^5\ M_\odot$, and for mass ratios $\eta\gtrsim10^{-3}$ for total mass $M=10^4\ M_\odot$.  For total mass $M=10^6\ M_\odot$, spin-spin coupling can be justifiably ignored for detection purposes.

One possible approach to developing IMRI waveforms would be to build hybrid waveforms that combine the post-Newtonian and perturbative waveforms.  There are at least two ways in which such waveforms can be developed.  Because EMRI waveforms have been expanded to very high post-Newtonian order \cite{Tagoshi:1996, Tanaka:1996}, one can develop a hybrid waveform family by adding the lowest-order in $\eta$ but high-order in $v/c$ terms to the standard post-Newtonian waveform.  On the other hand, one could add the $O(\eta^2)$ terms from the post-Newtonian expansion to the $O(\eta)$ EMRI waveform by equating observables such as the orbital frequency and frequency derivative between the two prescriptions. This trick has already been used to obtain the lowest order conservative correction to the EMRI phase evolution for a circular inspiral into a non-spinning black hole in~\cite{Babak:2007} and into a spinning black hole in \cite{Huerta:2009}, but can be readily extended to higher orders in $\eta$.  

We will not be able to test hybrid waveforms for faithfulness until advances in numerical relativity allow for direct simulations of intermediate-mass-ratio inspirals, or until progress on the self-force problem makes available approximate solutions at higher orders in $\eta$.  However, it may still be possible to find waveforms that effectively cover the space of ``true'' waveforms.  To do this, we will need to systematically check whether, for a given set of parameters, there is a hybrid waveform (not necessarily with the same intrinsic parameters) that has a high match to both post-Newtonian and EMRI waveforms.  Since ``true'' IMRI waveforms are likely to reside in the space between these approximations, such a hybrid family could be suitable for creating a bank of waveform templates for detection even if it is not faithful.

%Once such waveforms are developed, it will be necessary to determine whether they effectively cover the space of all waveforms in order to assess their suitability as detection templates.  To do this, we will need to systematically check whether, for a given set of parameters, there is a hybrid waveform (not necessarily with the same parameters) that has a high match to our model ``true'' waveform.

The requirements for parameter estimation are more strict than those for detection, since an effective family of waveforms that may be sufficient for detection can nevertheless yield significant statistical or 
systematic parameter estimation errors.  The statistical errors caused by the presence of noise can be analyzed with Fisher information matrix techniques or Markov Chain Monte Carlo searches (e.g., \cite{Wickham:2006}).  Meanwhile, systematic errors due to the use of approximate waveforms can be measured with a technique similar to the one proposed in \cite{CutlerVallisneri:2007}.  

The creation of IMRI waveforms, the analysis of parameter estimation accuracies, and the eventual development of new data analysis algorithms to enable searches for IMRIs with LISA will require a significant effort.  However, the potential benefits of IMRI observations make this effort worthwhile.  IMRIs may allow the first IMBH detections to be made.  They could add to our understanding of various astrophysical properties occurring in globular clusters and galactic nuclei.  Finally, they will serve as excellent probes of strong-field gravity, allowing us to measure whether central bodies are really Kerr black holes, and perhaps even making tests of general relativity possible \cite{Hughes:2006, Brown:2007}.  

\section*{Acknowledgments}
We would like to thank the Aspen Center for Physics, where this project was started.  We thank Stas Babak for useful discussions.  Ilya Mandel is partially supported by NASA ATP Grant NNX07AH22G to Northwestern University; his participation in the LISA Symposium was made possible by a
grant from the Symposium organizers and an International Travel Program grant from the AAS.  JG's work is supported by a Royal Society University Research Fellowship.

\section*{References}
\bibliographystyle{amsplain}
\bibliography{Mandel}

\providecommand{\bysame}{\leavevmode\hbox to3em{\hrulefill}\thinspace}
\providecommand{\MR}{\relax\ifhmode\unskip\space\fi MR }
\providecommand{\MRhref}[2]{%
  \href{http://www.ams.org/mathscinet-getitem?mr=#1}{#2}
}
\providecommand{\href}[2]{#2}
\begin{thebibliography}{10}

\bibitem{Amaro:2007}
P.~{Amaro-Seoane}, J.~R. {Gair}, M.~{Freitag}, M.~C. {Miller}, I.~{Mandel},
  C.~J. {Cutler}, and S.~{Babak}, \emph{{Intermediate and extreme mass-ratio
  inspirals -- astrophysics, science applications and detection using LISA}},
  Classical and Quantum Gravity \textbf{24} (2007), 113--+.

\bibitem{Arun:2005}
K.~G. Arun, Bala~R Iyer, B.~S. Sathyaprakash, and Pranesh~A Sundararajan,
  \emph{Parameter estimation of inspiralling compact binaries using 3.5
  post-{N}ewtonian gravitational wave phasing: The nonspinning case}, Physical
  Review D (Particles, Fields, Gravitation, and Cosmology) \textbf{71} (2005),
  no.~8, 084008.

\bibitem{MLDC2}
S.~{Babak} et~al., \emph{{Report on the second Mock LISA data challenge}},
  Classical and Quantum Gravity \textbf{25} (2008), no.~11, 114037--+.

\bibitem{Babak:2007}
S.~{Babak}, H.~{Fang}, J.~R. {Gair}, K.~{Glampedakis}, and S.~A. {Hughes},
  \emph{{``Kludge'' gravitational waveforms for a test-body orbiting a Kerr
  black hole}}, \prd \textbf{75} (2007), no.~2, 024005--+.

\bibitem{BarackCutler:2004}
L.~{Barack} and C.~{Cutler}, \emph{{LISA capture sources: Approximate
  waveforms, signal-to-noise ratios, and parameter estimation accuracy}}, \prd
  \textbf{69} (2004), no.~8, 082005--+.

\bibitem{Bender:1998}
P.L. {Bender} et~al., \emph{Lisa pre-phase a report; second edition}, Tech.
  Report MPQ233, 1998.

\bibitem{Colbert:2008}
C.~T. {Berghea}, K.~A. {Weaver}, E.~J.~M. {Colbert}, and T.~P. {Roberts},
  \emph{{Testing the Paradigm that Ultraluminous X-Ray Sources as a Class
  Represent Accreting Intermediate-Mass Black Holes}}, \apj \textbf{687}
  (2008), 471--487.

\bibitem{Blanchet:2006}
L.~{Blanchet}, \emph{{Gravitational Radiation from Post-Newtonian Sources and
  Inspiralling Compact Binaries}}, Living Reviews in Relativity \textbf{9}
  (2006), 4--+.

\bibitem{Brown:2007}
D.~A. {Brown}, J.~{Brink}, H.~{Fang}, J.~R. {Gair}, C.~{Li}, G.~{Lovelace},
  I.~{Mandel}, and K.~S. {Thorne}, \emph{{Prospects for Detection of
  Gravitational Waves from Intermediate-Mass-Ratio Inspirals}}, Physical Review
  Letters \textbf{99} (2007), no.~20, 201102--+.

\bibitem{CutlerVallisneri:2007}
C.~{Cutler} and M.~{Vallisneri}, \emph{{LISA detections of massive black hole
  inspirals: Parameter extraction errors due to inaccurate template
  waveforms}}, \prd \textbf{76} (2007), no.~10, 104018--+.

\bibitem{Ebisuzaki:2001}
T.~{Ebisuzaki}, J.~{Makino}, T.~G. {Tsuru}, Y.~{Funato}, S.~{Portegies Zwart},
  P.~{Hut}, S.~{McMillan}, S.~{Matsushita}, H.~{Matsumoto}, and R.~{Kawabe},
  \emph{{Missing Link Found? The ``Runaway'' Path to Supermassive Black
  Holes}}, \apjl \textbf{562} (2001), L19--L22.

\bibitem{Gralla:2008}
S.~E. {Gralla} and R.~M. {Wald}, \emph{{A rigorous derivation of gravitational
  self-force}}, Classical and Quantum Gravity \textbf{25} (2008), no.~20,
  205009--+.

\bibitem{Gultekin:2004}
K.~{G{\"u}ltekin}, M.~C. {Miller}, and D.~P. {Hamilton}, \emph{{Growth of
  Intermediate-Mass Black Holes in Globular Clusters}}, \apj \textbf{616}
  (2004), 221--230.

\bibitem{Gultekin:2006}
\bysame, \emph{{Three-Body Dynamics with Gravitational Wave Emission}}, \apj
  \textbf{640} (2006), 156--166.

\bibitem{Huerta:2009}
E.~A. {Huerta} and J.~R. {Gair}, \emph{{Influence of conservative corrections
  on parameter estimation for EMRIs}}, ArXiv e-prints (2008), 0812.4208.

\bibitem{Hughes:2006}
S.~A. {Hughes}, \emph{{(Sort of) Testing relativity with extreme mass ratio
  inspirals}}, Laser Interferometer Space Antenna: 6th International LISA
  Symposium (S.~M. {Merkovitz} and J.~C. {Livas}, eds.), American Institute of
  Physics Conference Series, vol. 873, November 2006, pp.~233--240.

\bibitem{Hurley:2007}
J.~R. {Hurley}, \emph{{Ratios of star cluster core and half-mass radii: a
  cautionary note on intermediate-mass black holes in star clusters}}, \mnras
  \textbf{379} (2007), 93--99.

\bibitem{Kidder:1993}
L.~E. {Kidder}, C.~M. {Will}, and A.~G. {Wiseman}, \emph{{Spin effects in the
  inspiral of coalescing compact binaries}}, \prd \textbf{47} (1993), 4183--+.

\bibitem{Mandel:2007rates}
I.~{Mandel}, D.~A. {Brown}, J.~R. {Gair}, and M.~C. {Miller}, \emph{{Rates and
  Characteristics of Intermediate Mass Ratio Inspirals Detectable by Advanced
  LIGO}}, \apj \textbf{681} (2008), 1431--1447.

\bibitem{Miller:2005}
M.~C. {Miller}, \emph{{Probing General Relativity with Mergers of Supermassive
  and Intermediate-Mass Black Holes}}, \apj \textbf{618} (2005), 426--431.

\bibitem{Miller:2009}
M.~C. {Miller}, \emph{{Intermediate-Mass Black Holes as LISA Sources}}, ArXiv
  e-prints (2008), 0812.3028; in this volume.

\bibitem{MillerColbert:2004}
M.~C. {Miller} and E.~J.~M. {Colbert}, \emph{{Intermediate-Mass Black Holes}},
  International Journal of Modern Physics D \textbf{13} (2004), 1--64.

\bibitem{MillerHamilton:2002b}
M.~C. {Miller} and D.~P. {Hamilton}, \emph{{Four-Body Effects in Globular
  Cluster Black Hole Coalescence}}, \apj \textbf{576} (2002), 894--898.

\bibitem{MillerHamilton:2002a}
\bysame, \emph{{Production of intermediate-mass black holes in globular
  clusters}}, \mnras \textbf{330} (2002), 232--240.

\bibitem{MouriTaniguchi:2002b}
H.~{Mouri} and Y.~{Taniguchi}, \emph{{Mass Segregation in Star Clusters:
  Analytic Estimation of the Timescale}}, \apj \textbf{580} (2002), 844--849.

\bibitem{MouriTaniguchi:2002a}
\bysame, \emph{{Runaway Merging of Black Holes: Analytical Constraint on the
  Timescale}}, \apjl \textbf{566} (2002), L17--L20.

\bibitem{Noyola:2008}
E.~{Noyola}, K.~{Gebhardt}, and M.~{Bergmann}, \emph{{Gemini and Hubble Space
  Telescope Evidence for an Intermediate-Mass Black Hole in {$\omega$}
  Centauri}}, \apj \textbf{676} (2008), 1008--1015.

\bibitem{OLeary:2006}
R.~M. {O'Leary}, F.~A. {Rasio}, J.~M. {Fregeau}, N.~{Ivanova}, and
  R.~{O'Shaughnessy}, \emph{{Binary Mergers and Growth of Black Holes in Dense
  Star Clusters}}, Astrophysical Journal \textbf{637} (2006), 937--951.

\bibitem{Poisson:2004}
E.~{Poisson}, \emph{{The Motion of Point Particles in Curved Spacetime}},
  Living Reviews in Relativity \textbf{7} (2004), 6--+.

\bibitem{PoissonWill:1995}
E.~{Poisson} and C.~M. {Will}, \emph{{Gravitational waves from inspiraling
  compact binaries: Parameter estimation using second-post-Newtonian
  waveforms}}, \prd \textbf{52} (1995), 848--855.

\bibitem{Pretorius:2005}
F.~{Pretorius}, \emph{{Evolution of Binary Black-Hole Spacetimes}}, Physical
  Review Letters \textbf{95} (2005), no.~12, 121101--+.

\bibitem{Sago:2008}
N.~{Sago}, L.~{Barack}, and S.~{Detweiler}, \emph{{Two approaches for the
  gravitational self-force in black hole spacetime: Comparison of numerical
  results}}, \prd \textbf{78} (2008), no.~12, 124024--+.

\bibitem{Scheel:2008}
M.~A. {Scheel}, M.~{Boyle}, T.~{Chu}, L.~E. {Kidder}, K.~D. {Matthews}, and
  H.~P. {Pfeiffer}, \emph{{High-accuracy waveforms for binary black hole
  inspiral, merger, and ringdown}}, \prd \textbf{79} (2009), no.~2, 024003--+.

\bibitem{Sesana:2008}
A.~{Sesana}, A.~{Vecchio}, M.~{Eracleous}, and S.~{Sigurdsson},
  \emph{{Observing white dwarfs orbiting massive black holes in the
  gravitational wave and electro-magnetic window}}, \mnras \textbf{391} (2008),
  718--726.

\bibitem{Smith:2009}
J.~R. {Smith} for~the LIGO Scientific~Collaboration, \emph{{The path to the
  enhanced and advanced LIGO gravitational-wave detectors}}, ArXiv e-prints
  (2009), 0902.0381.

\bibitem{Tagoshi:1996}
H.~{Tagoshi}, M.~{Shibata}, T.~{Tanaka}, and M.~{Sasaki}, \emph{{Post-Newtonian
  expansion of gravitational waves from a particle in circular orbit around a
  rotating black hole: Up to O($v^8$) beyond the quadrupole formula}}, \prd
  \textbf{54} (1996), 1439--1459.

\bibitem{Tanaka:1996}
T.~{Tanaka}, H.~{Tagoshi}, and M.~{Sasaki}, \emph{Gravitational waves by a
  particle in circular orbit around a schwarzschild black hole}, Prog. Theor.
  Phys. \textbf{96} (1996), 1087--1101.

\bibitem{Taniguchi:2000}
Y.~{Taniguchi}, Y.~{Shioya}, T.~G. {Tsuru}, and S.~{Ikeuchi}, \emph{{Formation
  of Intermediate-Mass Black Holes in Circumnuclear Regions of Galaxies}},
  \pasj \textbf{52} (2000), 533--537.

\bibitem{Trenti:2006}
M.~{Trenti}, \emph{{Dynamical evidence for intermediate mass black holes in old
  globular clusters}}, ArXiv Astrophysics e-prints (2006),
  arXiv:astro-ph/0612040.

\bibitem{Wickham:2006}
E.~D.~L. {Wickham}, A.~{Stroeer}, and A.~{Vecchio}, \emph{{A Markov chain Monte
  Carlo approach to the study of massive black hole binary systems with LISA}},
  Classical and Quantum Gravity \textbf{23} (2006), 819--+.

\end{thebibliography}

\end{document}